\documentclass[preprint,showpacs,preprintnumbers,amsmath,amssymb]{revtex4}

\usepackage{graphicx}
\usepackage{dcolumn}
\usepackage{bm}

\begin{document}


\title{Enhanced Raman scattering mediated by large wave-vector surface plasmon polaritons}

\author{T. L\'opez-R\'ios}
 \email{tomas.lopez-rios@grenoble.cnrs.fr}
\affiliation{
Institut N\'eel,CNRS et Universit\'e Joseph Fourier, BP 166, F-38042 Grenoble Cedex 9, France\\
}%

\date{\today}

\begin{abstract}
Surface enhanced Raman scattering mediated by surface plasmons polaritons with wave vector much larger than those of light is considered. The excitation of these polaritons and their efficient Raman scattering due to low group velocity and electric field localization at the surface is discussed for the case of two nearby surfaces. The Raman scattering depends not only on the intensity of the electric fields seen by the molecules but also on the surface plasmon densiy of states. 
\end{abstract}

\maketitle


Surface enhanced Raman scattering (SERS) is currently employed as an useful analytical tool. Nevertheless, a well established explanation of this effect is still lacking 
 \cite{1,2,3}. The different theories of SERS may be ranged into two types of quite different nature. The first ones, known as electromagnetic theories, assume that the Raman cross section of molecules is almost similar to those of isolated molecules and the Raman enhancement is caused by an amplification of the electromagnetic fields on the metallic surface where the molecules are located i. e. the Raman enhancement is directly related to the amplification of the local electromagnetic field seen by the molecule. The hypothesis most generally admitted is that this amplificaion is due to surface plasmons excitations.
 The second type, known as theories of the chemical effect, assume that the electronic interaction of the molecule with the metal surface leads to an increase of the Raman cross section similarly as in resonant Raman scattering.
 
In this communication I present a new mechanism that may account for observed  Raman enhancements. I will show that the Raman scattering cross section of molecules absorbed on a surface may be very large when Raman scattering is mediated by surface plasmons polaritons with large wavevector and low group velocity. 
I will show that the scattering of these quasiparticles by phonons strongly depends on their group velocity and may be much more efficient than the usual Raman scattering by light. This mechanism is formally and conceptually very similar to the enhancement of the resonant Brillouin or Raman scattering through excitons polaritons in crystals in the excitonic spectral region, a phenomena well established today  \cite{4,5}. In this communication I firstly discuss the Raman scattering mediated by surface plasmons polaritons and then I consider surface geometries which generate the peculiar surface polaritons giving rise to a very high Raman cross section.

Raman scattering mediated by surface plasmon-polaritons may be described in a rather similar way as resonant Raman scattering by excitons \cite{4} using a factorization of the different steps. I will assume  that the incoming photons scatter with the molecules in three steps schematically represented in Fig. 1, and summarized as follows:

\begin{enumerate}

\item an impinging photon with energy $\hbar \omega_i$ excites a surface polariton with energy $\hbar \omega_i$ and wave vector $k_i$. The probability (or efficiency) of such conversion is $T_i$.  

\item the surface polariton ($\hbar \omega_i$ , $k_i$) subsequently scatter  with a molecule creating a phonon of energy $\hbar \Omega$ (for simplicity  only the Stokes process is considered) and a surface plasmon polariton of energy $\hbar \omega_s = \hbar \omega_i - \hbar \Omega$ and wave vector $k_s$. Note that $k_i$ and $k_s$ are wavevectors both on the surface and direction according to the momentum conservation.

\item the  surface plasmon polariton ($\hbar \omega_i - \hbar\Omega , k_s$) is converted back into a photon with an efficiency $T_s$, and finally collected at the detector.
\end{enumerate}

\begin{figure}
\includegraphics[scale=0.7]{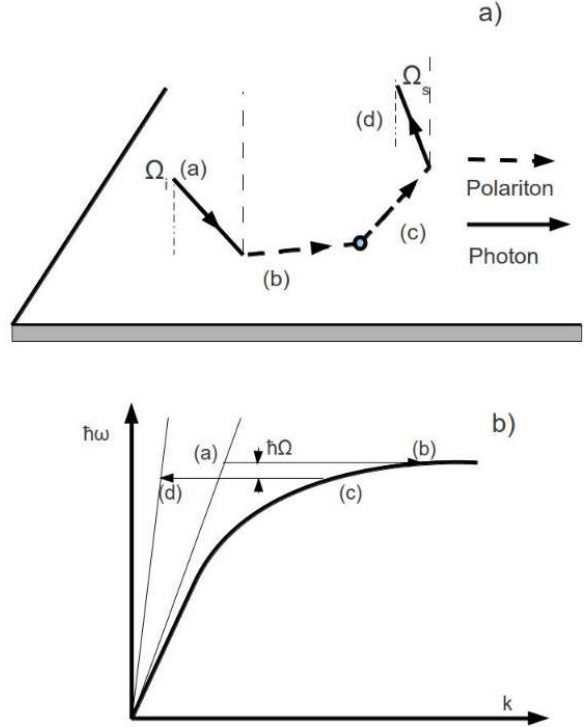}
\caption{\label{fig:epsart}  a) Schematic representation of Raman scattering by surface polaritons:  photon \textit{(a)} impinging on the surface with an angle $\Omega_i$  generating  a surface polariton with wavevector $k_i$ \textit{(b)} which scatters with a molecule into a polariton with wavevector $k_s$ \textit{(c)} that converts into a photon \textit{(d)} leaving the surface at an angle $\Omega_s$. 
b) Dispersion of surface plasmons and light. Polaritons are indicated in \textit{(b)} and \textit{(c)} and photons in \textit{(a)} and \textit{(d)}. }
\end{figure}

The probability for a polariton $\vert k_i \rangle$ to scatter into a polariton  $\vert k_s \rangle$ by the interaction with a potential $V$ with the molecule is given, to the first order approximation, by the Fermi transition rate golden rule: 

\begin{eqnarray}
\left(\frac{dP^{Pol}}{dt} \right)_{k_i \rightarrow k_s}  = \frac{2 \pi}{\hbar} \vert \left< k_i \vert V \vert k_s \right> \vert^2 \delta \left(\hbar\omega_i - \hbar\omega_s - \hbar\Omega \right) \nonumber
\end{eqnarray}

The overall scattering probability account for all the possible final states $k_s$ in the $k-$plane. The differential element on the two dimensional  $k-$surface is $ds = k_s \, d\alpha \, dk_s$  and the number of states between $k$ and $k+dk$ is $k_s \, d\alpha \, dk_s / (2\pi)^2$. Integrating around all the possible directions in the plane and including $dk = d\omega /(d  \omega / d  k )$ we find that the overall scattering probability is: 

\begin{eqnarray}
\frac{dP^{Pol}}{dt}  = \frac{2 \pi}{\hbar} \vert \left< k_i \vert V \vert k_s \right> \vert^2 \rho(\omega) \nonumber
\end{eqnarray}
where $\rho(\omega) = k_s/(2\pi \hbar v_{gs})$ is the surface polariton density  of states and $v_{gs} = (d  \omega_s /d  k)$ the group velocity of the scattered surface polaritons. Without damping, $v_{gs}$ corresponds to the energy velocity. $P^{Pol} = (dP^{Pol}/dt ) \tau$, where $\tau$ is the life time of the excited polaritons that in the case of long wave vector can be approximated by \cite{6}:

\begin{eqnarray}
\tau = \frac{1}{2\epsilon^{\prime\prime}}\frac{d  \epsilon^\prime}{d  \omega}
\end{eqnarray}

An estimation with the dielectric constants \cite{7} gives for the damping for $\omega \sim \omega_s$, $\hbar/\tau \sim $ 0.1 eV in agreement with low energy loss experiments \cite{8}. The polariton life time is of about 4$\cdot$10$^{-14}$ s. \\

Finally, in the global process, the scattering efficiency for a photon $\hbar \omega_i$ striking the surface in the $\Omega_i$ direction to generate a photon $\hbar \omega_s$ in the solid angle $d\Omega_s$ around $\Omega_s$ is

\begin{eqnarray}
\frac{dP}{d\Omega_s} = T_i (\Omega_i) T_s (\Omega_s) \frac{k_s}{\hbar^2v_{gs}} \vert \left< k_i \vert V \vert k_s \right> \vert^2 \tau 
\end{eqnarray}
with $\tau$ given by Eq. 1. It is explicitly indicated that $T_i (\Omega_i)$ is the probability to excite a surface polariton by an incoming photon in the direction $\Omega_i$ and $T_s(\Omega_s )$ is the probability for a surface polariton to generate a photon in the direction $\Omega_s$. Photons are collected in the solid angle  $d\Omega_s$ around $\Omega_s$. Insofar as the surface polaritons propagation length is microscopic in our case, the outgoing photons are collected in the experiment at the same point as the impinging one. $T_i (\Omega_i)$ depends on the surface polariton density of states $\rho (\omega_i ) = k_i/2\pi\hbar v_{gi}$ where  $v_{gi} = (d  \omega_i / d  k )$ is the group velocity of the excited surface polaritons. We can admit that the probability that a photon converts into a polariton is proportional to the number of allowed polaritos, i.e., the polariton density od states and thus $dP/d\Omega_s \sim 1/v_{gi}v_{gs}$. As the group velocity can be small this factor increases the Raman scattering. This is a first possible mechanism of Raman enhancement. \\

Low group velocity ($v_g/c \ll 1)$ has been measured for exciton polaritons \cite{9} and enhancement by several orders of magnitude of the Raman scattering of semiconductors due to this velocity factor are well documented in the literature \cite{4,5}. Enhancement of Raman scattering by slow group velocity in non-excitonic media has already been evoked \cite{10} and experimentally observed \cite{11} in photonic crystals for energies approaching the band edge. Stimulated Raman scattering enhanced by the same mechanism  was predicted in optical guides \cite{12}. 
These are some examples of the numerous slow light situations giving rise to interesting phenomena \cite{13} also observed in other kinds of waves like ultrasound propagating on elastic plates \cite{14}.\\

The second possible mechanism of Raman enhancement is the term $\vert \left< k_i \vert V \vert k_s \right> \vert^2$ in Eq. 2. In the classical dipolar approximation, this term is proportional to the intensity of the incident and scattered fields seen by the molecule (electrics fields at $z$ =0). According to Eq. 5 of Appendix, the electric field intensity increases linearly with $k$ approaching the frequency $\omega_{sp}$ for which $\epsilon \rightarrow -1$, $k \rightarrow \infty$ and $v_g \rightarrow 0$. 
In summary  $dP/d\Omega_s \sim k_i k_s^2/ v_{gi} v_{gs}$  may become extremely large for photons with energies close to $\omega_{sp}$.
Nevertheless for real metals with flat surfaces, damping prevents to have surface plasmons with $v_{gi}, v_{gs} \ll c$  and high fields at the surface \cite{15}.
In Raman scattering mediated by surface plasmons polaritons excited by attenuated total reflection (ATR) as in ref. \cite{16}, $v_{gi} \approx v_{gs} \approx c$ and the enhancement is due to the modest electromagnetic fields amplification at the surface. \\

Today is generally admitted that high electromagnetic fields in gaps or cavities of nanometric or sub-nanometric dimensions leads to SERS  \cite{17,18,19,20}. In the case of metallic particles the electromagnetic enhancement increases 
on reducing the distance between the particles \cite{21}.
An extremely important point of our argumentation about Raman enhancement is that for more complex surfaces than a flat surface, long wave vector surface plasmons polaritons can be spontaneously generated  simply by shining the sample with light. \\ 

It was recently theoretically demonstrated \cite{22} and experimentally observed \cite{23} that rectangular groves of nanometric dimension on otherwise flat surfaces may strongly absorb the impinging light by the excitation of standing surface plasmons polaritons  waves in the cavities. In such a case the conversion of photons to long wave vector surface plasmons polaritons is very efficient. The Le Perchec \textit{et al} results \cite{22} are extremity important in the present discussion because they demonstrate that light may generate electronic surface plasmons polaritons 
with high electric field and small magnetic fields. 
More  recently it was demonstrated that this  type of polaritons exists for nano-metric cavities of a variety of shapes \cite{24}.\\

Moreover, some other indications lead to think that plasmons with extremely long wavevector are excited when  SERS occurs. Namely, the inelastic background assigned to a very rapid recombination of electron hole pairs generated  by the impinging light are necessarily created by quasiparticles with momentum of the order of $k_F$.\\

One feature of SERS is that the laser light is largely absorbed by the samples in spite of the fact that the metal itself must be very reflecting. Optical absorptions are observed, for instance in colloids, discontinuous films or deposits on a cold substrate. 
Differential reflectivity measurements of Ag films deposited on a cold substrate have shown that the  surface absorption follows the SERS excitation spectra  \cite{25}. These experiments show that many surface plasmons can be excited in rough surfaces but only those with the electric fields strongly localized at the surface contribute to SERS.\\

In a prophetic paper written almost one century ago about the anomalous optical absorption of Na condensed on cold substrates R. W. Wood underlined \cite{26}:  "the spaces or cavities between the sodium crystals act as traps for radiation of definite frequencies". It is well known today that this spaces or cavities have sub-nanometric dimensions. \\

The left panel of Fig. 2a schematically represents notches or grain boundaries of a metallic sample and the right side the spaces between two closed particles like coalesced colloids or grains of discontinuous films. Fig. 2b shows two geometries to model such features. The electromagnetic fields for the geometries represented in this figure were extensively investigated \cite{27,28,29,30,31}). In both cases (closed and open cavities) resonances are generated by stationary waves built up by modes of two infinite metallic planes with antisymmetric charge distribution. Surface charges on the cavities for a first order resonance are indicated in the figure. The electric field is nil at the bottom of the closed cavity and at the middle of the channel for the open cavities, similarly to sound intensity in pipe resonances. Actually, Rayleigh in a posthumous publication concerning acoustic resonances in a  perforated wall \cite{32} wrote that sound may behave as light in cavities of  alkali films studied  by Wood \cite{26}. 
To discuss polaritons in the geometries indicated in Fig. 2b I consider the well known electromagnetic modes of two interacting metallic surfaces as indicated in the inset of Fig. 3. \\

\begin{figure}
\includegraphics[scale=0.6]{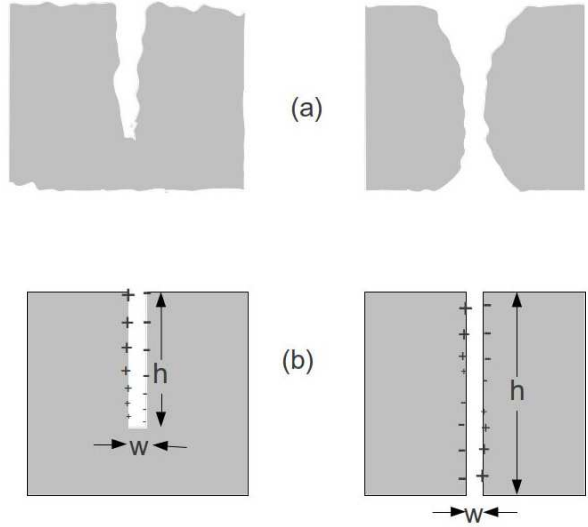}
\caption{\label{fig:epsart} a) Representation of geometries with nanometric dimensions giving rise to SERS : a notch in a metal surface and a interstice between two closed particles. b) Idealization of the geometries above with flat surfaces of length $h$ separated by a distance w. In the figure the surface charges for standing waves resonances are qualitatively indicated.}
\end{figure}

The surface plasmon dispersion relations of two flat surfaces of identical metals (of dielectric constant $\epsilon \equiv \epsilon^\prime + i \epsilon ^{\prime \prime}$) separated by a distance w can be found in ref. \cite{33} and in the case relevant for us with $k \gg \omega/c$ are given by:

\begin{eqnarray}
\frac{\epsilon +1}{\epsilon - 1} = \pm e^{-k\mathrm{w}}
\end{eqnarray}
where $k \equiv k^\prime + i k^{\prime \prime}$ is the complex wavevector parallel to the surface. 
The "+" solution is the antisymmetric mode which exists for frequencies $< \omega_{sp}$  that is the only one I will consider. The real and imaginary part of $k$ normalized to w are shown in Fig. 3 for silver taking for $\epsilon $ values given in ref. \cite{7}. It is important to underline that \textit{for a given frequency, $k$ can be larger, the smaller the gap while having a relative small imaginary part. This is an important property and an essential difference with surface plasmons of a single surface}. It can be shown that the fields at the surface are given by the same expression as that for a single surface i.e by Eq. A5 (see Appendix) but with $k$ given by Eq. 3. Very clearly the electric field become very high approaching $\omega_{sp}$. In the case that w is extremely small ($k\mathrm{w} \ll 1$ with  $k \gg \omega/c$) we have from Eq. 3 that $k \approx 2/\mathrm{w} \vert \epsilon-1 \vert$. In this limit resonances of cavities of Fig. 3b occurs for $h$ much smaller than the wavelength of light. \\

In the spectral region with $\epsilon^{\prime\prime} \ll \epsilon^\prime$, and from Eq. 3, the group velocity is given, to a good approximation, by 

\begin{eqnarray}
\frac{1}{c}\frac{d  \omega}{d  k} \sim \frac{1}{2} \mathrm{w} (\epsilon^\prime - 1) (\epsilon^\prime + 1)\left(\frac{d  \epsilon^\prime}{d  \omega} \right)^{-1} \nonumber
\end{eqnarray}

The dielectric constant of Ag can be represented as the sum of two contributions : the $s-p$ near free electrons, $\epsilon^f = 1-\omega_p^2/\omega^2$, and the bounded  $d-$band electrons contribution, $\epsilon^b$ \cite{34}. 
Neglecting the dispersion in the $d-$electron contribution  
we have:

\begin{eqnarray}
\frac{1}{c}\frac{d  \omega}{d  k} \sim \mathrm{w} \frac{\omega}{4c} (\epsilon^\prime - 1)(\epsilon^\prime + 1) \left(\frac{\omega_{p}^2}{\omega^2} \right)^{-1}
\end{eqnarray}

Fig. 3 shows  $(d  \omega/d  k)/c$ for $\mathrm{w} =$ 1 nm computed with Eq. 4 and $\hbar \omega_p = $9 eV. \\

In conclusion, for sufficiently small w polaritons with long wave vectors, high electric fields at the surface and small group velocity exist in a rather large spectral region. More important, for these modes the real part of the wave vector can be large with a relatively small imaginary part provide that w is sufficiently small. \\ 

\begin{figure}
\includegraphics[scale=0.55]{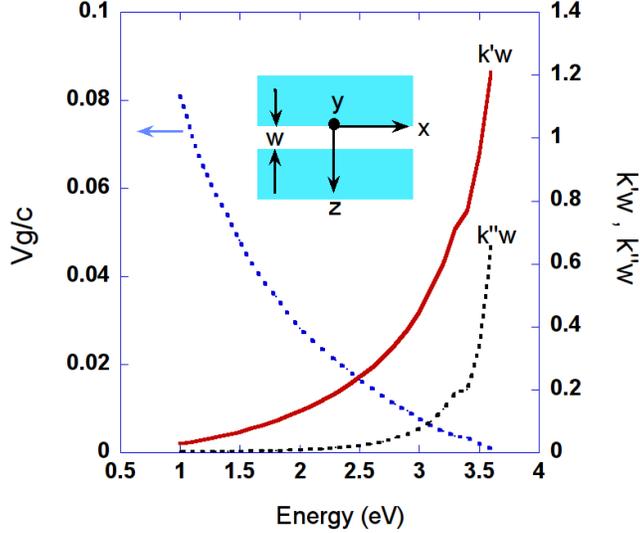}
\caption{\label{fig:epsart} Group velocity $v_g$ of surface plasmons of two silver surfaces separated by w=1 nm and $k^\prime$w and $k^{\prime\prime}$w for two closed surfaces with the approximation $k \gg \omega/c $.}
\end{figure}

The factor $dP/d\Omega_s \sim k_i k_s^2/ v_{gi} v_{gs}$ can be very high for plasmons of the geometry indicated in the inset of Fig. 3.
Around 3.5 eV this factor becomes very high but its accuracy may be questioned due the increasing damping and the validity of the local approximation. Nevertheless, an estimation of this factor at 2 eV, where the approximation is safer, shows that it is about 3$\cdot$10$^6$ larger than that for Raman scattering mediated by surface plasmons polaritons of the same energy 
in an ATR experiment ($v_{gi} \approx  v_{gs} \approx c$ and $k_i \approx ks \approx  \omega/c$). An enhancement of 4$\cdot$10$^4$ was determined for Raman scattering mediated by surface plasmons of a flat surface excited by photons of 2.4 eV in an ATR configuration  \cite{16}). A realistic Raman enhancement must be determined from the electric fields and velocities of polaritons for each specific geometry. \\

It is noteworthy to remark that first order Raman scattering by surface plasmon polaritons may involve phonons with a noticeable momentum. This is not the case for Raman scattering with light. This is particularly true for polaritons with frequencies close to $\omega_{sp}$ for which momentum can be conserved in the scattering process even for phonons with large  momentum. This may explain some intriguing differences concerning the Raman frequencies observed in SERS with respect to the molecules in solutions, sometimes attributed to molecular distortion by chemisorption. 
It is also important to point out that the hypothesis presented herein also applies to Brillouin scattering. Consequently Brillouin scattering enhancement of bulk and surface phonons should be observed in SERS, as well. This is probably the case as evidence structures observed on the low frequency region of the inelastic background. In particular, it was observed, for silver deposited on cold substrates, features in the SERS background which corresponds to the phonon density of states \cite{35}. These structures can be understood as enhanced Brillouin scattering at metal surfaces.  \\               
 
All the features of SERS are consistent with the ideas presented herein. For instance the extremely large surface sensitivity of SERS \cite{36}, unusual for an optical phenomena, can be easily understood by the nature of the long wave vector surface polaritons. The presence of hot spots (or active sites) may also be understood as the few places on the surface where the photon polariton conversion occurs.  \\


This paper shows that the Raman scattering depends not only on the intensity of the electric fields seen by the molecule but also on the surface plasmon density of states. Raman scattering mediated by surface polaritons is enhanced because : \textit{(i)} The density of states of these surface polaritons may be very large.  \textit{(ii)} Long wave vector surface polaritons have larger electric fields at the surface than photons of the same energy. Moreover the Raman selection rules are also different than for the usual spontaneous Raman scattering. Plasmons or phonon polaritons on surfaces may propagate with a velocity much smaller than those of light and have a longer interaction time with molecules at surface. For some surface spectroscopies (beyond SERS), it is important to consider that short wave-length polaritons may interact with molecules more efficiently than photons do. \\ 

I am indebted firstly to A. Barbara and P. Qu\'emerais for multiple discussions during the past years, and to P.S. Apell, J.E. Lorenzo and A. Wirgin for a critical reading of the manuscript. Finally I thanks A. Liebsch, C. L\'opez-R\'ios and D. Mayou for discussions on the issues considered in this paper. \\

\appendix
\section{\label{sec:level2}ELECTRIC FIELDS AND GROUP VELOCITY OF SURFACE PLASMONS \protect\\}

Let consider a flat surface (in the $x-y$ plane) between vacuum (medium 1) and a metal (medium 2) of dielectric constant $\epsilon$ with a surface plasmon of wave vector $\mathbf{k}$ in the $x-$direction and the magnetic field in the $y-$direction given by 

\begin{eqnarray}
(\mathbf{H_1})_y &=& H e^{ikx} e^{q_1z} e^{-i\omega t} \nonumber \\
(\mathbf{H_2})_y &=& H e^{ikx} e^{-q_2z} e^{-i\omega t}
\end{eqnarray}

with $q_1 = \sqrt{k^2 - (\omega /c)^2}$ and $q_2 = \sqrt{k^2 - (\omega /c)^2 \epsilon}$. The electric fields are given by the relation  $\nabla\times \mathbf{H} = -i \omega \epsilon \epsilon_0 \mathbf{E}$.

\begin{eqnarray}
(\mathbf{E_1})_x & = &-i H (q_1/\omega\epsilon_0) e^{ikx} e^{q_1z} e^{-i\omega t} \nonumber \\ (\mathbf{E_1})_z & = &- H (k/\omega\epsilon_0) e^{ikx} e^{q_1z} e^{-i\omega t} \nonumber  \\
(\mathbf{E_2})_x & = & i H (q_2/\omega\epsilon_0)e^{ikx} e^{-q_2z} e^{-i\omega t} \nonumber \\ (\mathbf{E_2})_z & = & - H (k/\omega\epsilon_0)e^{ikx} e^{-q_2z} e^{-i\omega t}
\end{eqnarray}


The dispersion relation of this wave is obtained from the continuity of the components of the electric and magnetic fields parallel to the surface :

\begin{eqnarray}
k = \frac{\omega}{c} \sqrt{\frac{\epsilon}{\epsilon+1}}
\end{eqnarray}
Without absorption 
the group velocity $v_g = d  \omega / d  k$ obtained from Eq. A3 is given by :
\begin{eqnarray}
v_g/c = \frac{\sqrt{\vert \epsilon \vert} {\vert \epsilon + 1 \vert}^{3/2}} { \epsilon (\epsilon+1) + \frac{1}{2} \omega \frac{d  \epsilon }{d  \omega}}
\end{eqnarray}
where $c$ is the light velocity in vacuum. At the surface plasmons of frequency $\omega_{sp}$, $\epsilon \rightarrow -1$, $ k \rightarrow \infty$ and $v_g \rightarrow 0$.  

Equations A1 and A2 show that for $k \gg (\omega/c)\sqrt{\epsilon}$ the electric field is much larger than the magnetic field and decay out of the surface over distances $\sim 1/k$. 

The density of electromagnetic radiation for a non magnetic media ($\mu$ = 1) with real dielectric constant is given  \cite{35} by :

\begin{eqnarray}
\mathrm{U} =  \frac{1}{4} \left(\epsilon_0 \frac{d (\omega \epsilon) }{d  \omega} \mathrm{E}^2 + \mu_0 \mathrm{H}^2 \right) \nonumber
\end{eqnarray}
where $\mathrm{U}$ is in MKS units and E$^2$ and H$^2$ are the intensity of the electric and magnetic fields, respectively. The time averaged energy density in media 1 over a surface $S$ is obtained by integrating in $z$ from $-\infty$ to 0.

\begin{eqnarray}
U_1 =  \frac{S}{8q_1} \left(\epsilon_0 E^2 + \mu_0 H^2 \right) =  \frac{S E^2 \epsilon_0}{4q_1}  \frac{\vert \epsilon \vert}{\vert \epsilon -1 \vert}\nonumber
\end{eqnarray}
where $H^2$ and $E^2=H^2(q_1^2+k^2)/\omega^2\epsilon_0^2$ are the intensities of the magnetic and the electric fields in the vacuum at the surface ($z$=0). The corresponding energy density in media 2 is:

\begin{eqnarray}
U_2 =  \frac{SE^2\epsilon_0}{4\vert \epsilon \vert q_1} \left( \frac{1}{\vert \epsilon -1 \vert}+ \frac{\omega}{2\vert \epsilon \vert} \frac{d  \epsilon }{d  \omega} \right) \nonumber
\end{eqnarray}

And the overall density of the energy in both media  :

\begin{eqnarray}
U =  \frac{SE^2\epsilon_0}{4\vert \epsilon \vert^2 q_1} \left( \vert \epsilon \vert\vert \epsilon +1 \vert + \frac{\omega}{2} \frac{d  \epsilon }{d  \omega} \right) \nonumber
\end{eqnarray}

In the case that a photon of energy $\hbar \omega$ converts into a polariton with the energy  distributed over a surface $S$, the electric field intensity of the  polariton at the surface will be :

\begin{eqnarray}
E^2 \approx  \frac{4 \hbar \omega \epsilon^2 q_1}{S\epsilon_0}  \left( \vert \epsilon \vert\vert \epsilon +1 \vert + \frac{\omega}{2} \frac{d \epsilon}{d \omega}\right) 
\end{eqnarray}
For long wave vectors ($k \gg (\omega/c)\sqrt{\vert \epsilon \vert}$), $E^2 \sim k$.

\end{document}